\documentclass[%
 reprint,
superscriptaddress,
 amsmath,amssymb,
 aps,
]{revtex4-1}

\usepackage{bbold}
\usepackage{mathptmx}
\usepackage{subfig}
\usepackage{psfrag,graphicx}
\usepackage{dcolumn}
\usepackage{amsmath,amssymb}
\usepackage{bm}
\usepackage{color}
\usepackage{latexsym}
\usepackage{epstopdf}
\usepackage{color}
\usepackage[english]{babel}
\usepackage{latexsym}
\usepackage{psfrag,graphicx}
\usepackage{amsmath}
\usepackage{amssymb}
\usepackage{amsfonts}
\usepackage{bm}
\usepackage{natbib}
\usepackage{epstopdf}
\usepackage{appendix}
\usepackage{rotating}
\usepackage[english]{babel}
\usepackage{aeguill}
\usepackage{ulem}
\usepackage[justification=justified]{caption}

\definecolor{mygrey}{gray}{0.35}
\definecolor{myblue}{rgb}{0.2,0.2,0.8}
\definecolor{myzard}{cmyk}{0,0,0.05,0}
\definecolor{mywhite}{rgb}{1,1,1}
\definecolor{mywhite}{rgb}{1,1,1}
\definecolor{myred}{rgb}{1,0.,0.3}

\usepackage[colorlinks=true,citecolor=myblue,linkcolor=myred]{hyperref}

\def\ba{\begin{align}}
\def\enda{\end{align}}
\def\bi{\begin{itemize}}
\def\ei{\end{itemize}}

\def\be{\begin{equation}}
\def\ee{\end{equation}}
\def\bea{\begin{eqnarray}}
\def\eea{\end{eqnarray}}
\def\bse{\begin{subequations}}
\def\ese{\end{subequations}}

\newcommand{\ket}[1]{|{#1}\rangle}                       % ket
\newcommand{\bra}[1]{\langle {#1}|}                      % bra
\newcommand{\average}[1]{\langle {#1} \rangle}           % media < >

\newcommand{\Ignore}[1]{ }

\def\i{\text{i}}

\begin{document}

\preprint{APS/123-QED}

%\title{Greenberger-Horne-Zeilinger state generation in qubit-chains via a single Landau-Majorana-St\"uckelberg-Zener pulse}
\title{Greenberger-Horne-Zeilinger state generation in qubit-chains via a single $\pi/2$-pulse}

\author{R. Grimaudo}
\address{Dipartimento di Fisica e Chimica ``Emilio Segr\`{e}",
Universit\`{a} degli Studi di Palermo, viale delle Scienze, Ed. 18, I-90128, Palermo, Italy}

\author{N. V. Vitanov}
\address{Department of Physics, St. Kliment Ohridski University of Sofia, 1164 Sofia, Bulgaria}

\author{A. S. M. de Castro}
\address{Universidade Estadual de Ponta Grossa, Departamento de F\'{\i}sica, CEP 84030-900, Ponta Grossa, PR, Brazil}

\author{D. Valenti}
\address{Dipartimento di Fisica e Chimica ``Emilio Segr\`{e}",
Universit\`{a} degli Studi di Palermo, viale delle Scienze, Ed. 18, I-90128, Palermo, Italy}

\author{A. Messina}
%\address{INFN, Sezione di Catania, I-95123 Catania, Italy}
\address{ Dipartimento di Matematica ed Informatica, Universit\`a degli Studi di Palermo, Via Archirafi 34, I-90123 Palermo, Italy}

\date{\today}

\begin{abstract}
A protocol for generating Greenberger-Horne-Zeilinger states in a system of $N$ coupled qubits  is proposed.
The Hamiltonian model assumes $N$-wise interactions between the $N$ qubits and the presence of a controllable time-dependent field acting upon one spin only.
Implementing such a scenario is in the experimental reach.
The dynamical problem is exactly solved thanks to the symmetries of the Hamiltonian model. 
The possibility of generating GHZ states under both adiabatic and non-adiabatic conditions is shown and discussed in detail. 
\end{abstract}



\maketitle

%\section{Introduction}
\textit{Introduction.}
In the last two decades a growing attention has been paid on Greenberger-Horne-Zeilinger (GHZ) states \cite{GHZ} and their generation \cite{Zhao,Li,Carvacho,Bishop,Su}.
These states, as well as $W$-states and Dicke-states, exhibit many-body or multi-partite entanglement, that is quantum correlations involving all the parts which a system consists of \cite{Amico}.

Entanglement plays a central role in quantum mechanics as an essential resource for several applications in quantum
information \cite{Terthal}, quantum communication \cite{Kimble}, quantum metrology \cite{Pezze, Carollo1}, quantum topology \cite{Carollo2} and also provides the possibility to test quantum mechanics against local hidden theory \cite{GHSZ}.
Further, it is at the basis of far-reaching discoveries such as quantum teleportation \cite{Bennett93},  quantum dense coding \cite{Bennett92}, quantum computation \cite{Steane}, quantum cryptography \cite{Ekert} and quantum fingerprinting \cite{Buhrman}.

Multipartite entangled states are of great interest for all these fields.
They are exploited to speed up computations \cite{Grover}, secure private communications \cite{Ren}, and to overcome the standard quantum limit \cite{Riedel}.
Among all the numerous examples of multipartite entangled states, the GHZ states play a special role in quantum information \cite{Song} and provide a core resource for applications in quantum metrology \cite{Omran} and quantum error correction \cite{Nielsen-Chuang}. 
Moreover, it is remarkable the pivotal place occupied by these states to test fundamental aspects of quantum mechanics \cite{GHSZ}.

The increased awareness of such a broad applicability has undoubtedly stimulated the emergence of numerous theoretical investigations focused on the search for realistic experimental protocols effectively exploitable in laboratory for the generation of GHZ states.
Notwithstanding, creating GHZ states of a system of $N$-qubits in a robust manner is still a current challenging and topical problem \cite{Zhao,Li,Carvacho,Bishop,Su}.
Many schemes have been proposed to produce such multipartite states via a single- or multi-step process \cite{Saffman}.
GHZ states have been previously experimentally realized through nuclear spin systems \cite{Neumann}, optical photons \cite{Wang}, trapped ions \cite{Friis}, and superconducting quantum circuits \cite{DiCarlo}.

As the number of qubits of the physical system increases, the generation of GHZ states comes up against technical difficulties that prevent the size of these multipartite states from being controllable at will.
In the past few years, the highest numbers of qubits that have been effectively entangled in a GHZ state include 14 trapped ions \cite{Sackett}, 18 state-of-the-art
photon qubits \cite{Zhong}, and 12 superconducting qubits \cite{Gong}.
It is then a matter of fact that so far the practical realization of experimental schemes for generating GHZ states often progressively loses its effectiveness when the number of qubits exceeds ten. 

In this work we propose a novel and straightforward procedure for constructing GHZ states and, more generally, many-body entangled states of $N$-qubit systems.
The experimental scheme we propose in this paper is based  on a new exactly solvable time-dependent $N$-qubit model \cite{GLSM}.
%The seminal idea reported in this reference is here exploited and developed adopting an original approach.
%Reference \cite{GLSM} has in fact a more speculative character and its scope is mainly centred on the presentation of a time-dependent many-body spin model more complex than that assumed in this paper.
%The rich dynamical properties of the model reported in Ref. \cite{GLSM} have been investigated mainly focusing on the specificity of the engineered $N$-wise coupling between the $N$ qubits.
Reference \cite{GLSM} has a speculative character, mainly focusing on the specificity of the engineered $N$-wise coupling between the $N$ qubits.
The time-dependent model we use here, instead, is firmly anchored to the two most prominent physical systems suitable for quantum information and computation: trapped ions and superconducting qubits.
%In this paper we make use of a comparatively simpler  time-dependent model, tailored in such a way to be firmly anchored to two crucial physical systems suitable for quantum information and computation: trapped ions and superconducting qubits.
It has been devised, indeed, having in mind: first, the well consolidated protocols for engineering $N$-body interactions involving all qubits of the system ($N$-wise interaction) both in trapped ion \cite{Barreiro, Muller} and superconducting qubit systems \cite{MezzacapoPRL113}; second, the ability of applying effective time-dependent fields, in principle at will, on just one qubit by only performing single-qubit operations in the case of superconducting qubits \cite{MezzacapoPRL113} and through the Scanning Tunneling Microscopy (STM) technique in the case of trapped ions \cite{Yan,Bryant,Tao}.

We show that, after generating the $N$-wise interaction, it is possible to generate GHZ states and, more in general, maximally entangled states of qubit systems by applying a single Landau-Majorana-St\"uckelber-Zener (LMSZ) pulse (a linear ramp) \cite{LMSZ} on just one qubit (ancilla).
Our analysis transparently reveals the peculiarity of the $N$-wise interaction which works as a `quantum bus'.
It is capable of coherently reverberating the dynamic beahviour of the ancilla qubit on all the other qubits of the system.

%Such a peculiar $N$-wise coupling characterizing our model is, of course, alien to physical context like nuclear, atomic, and molecular physics.
%However, such $N$-spin Hamiltonian models turn out to be remarkably exploitable for quantum information applications \cite{GLSM}, to treat and study fermion lattice models \cite{Casanova} and to describe better physical features and dynamic aspects of complex systems \cite{Zylberberg}.
The peculiar $N$-wise coupling we consider is, of course, alien to physical context like nuclear, atomic, and molecular physics.
However, such $N$-spin Hamiltonian models turn out to be remarkably exploitable to treat fermion lattice models \cite{Casanova} and to describe better physical features of complex systems \cite{Zylberberg}.

%The paper is organized as follows...

%\section{The Model and its Symmetries}
\textit{The Model.}
The physical system we are about to investigate in this paper consists in a spin-chain composed of $N$ distinguishable spin-1/2's interacting through an $N$-wise interaction term which is characterized by the coupling strength parameter $\gamma_x$.
We assume that it is described by the following Hamiltonan model: 
\begin{equation} \label{Model}
H=\hbar\omega_1(t) \hat{\sigma}_1^z + \gamma_x ~ \bigotimes_{k=1}^N \hat{\sigma}_k^x,
\end{equation}
where $\hat{\sigma}_k^x$ $\hat{\sigma}_k^y$ and $\hat{\sigma}_k^z$ are the standard Pauli matrices of the $k$-th spin in the chain.
In general \cite{GLSM} $N$-wise means that the interaction among the $N$ qubits may be represented as an $N$-degree homogeneous multilinear polynomial in the 3$N$ dynamical variables of all $N$ qubits.
The first term in the Hamiltonian expression, instead, is due to the application of a field, in general time-dependent, on the first spin of the chain which induces the energy separation $\hbar \omega_1 (t)$.

The model under scrutiny is of significant interest in quantum information.
The peculiar $N$-wise interaction we assume here, can be in fact implemented in the two main physical systems of paramount applicative interest for quantum computation: superconducting qubit arrays \cite{MezzacapoPRL113} and trapped ion systems \cite{Barreiro, Muller}.
In the first case, fast multiqubit interactions are realized by tuning the transmon-resonator couplings through the modulation of magnetic fluxes  \cite{MezzacapoPRL113}.
In the second case, instead, the main resource of the simulation technique are multi-ion M{\o}lmer-S{\o}rensen \cite{MS} gate operations, based on the application of a bichromatic laser field to the ions \cite{Muller}.

As far as the effective time-dependent field is concerned, in the superconduncting-qubit scenario it can be locally generated on one spin only, chosen at will in the chain, through single-qubit gates.
Precisely, by detuning the respective qubit from its idle frequency by an amount $\delta$, it is possible to generate field strength of $2\pi\delta$ \cite{Salathe}.
In the trapped ions/atoms scenario, instead, the application of a time-dependent field on a single spin of the chain can be realized through the Scanning Tunneling Microscopy (STM) technique or by microwave approach \cite{Yan,Bryant,Tao}.
The latter, for example, allows to produce an LMSZ ramp in a time window of $20 ps$ on a single spin of a chain \cite{Sivkov}.
The free Hamiltonian of a prefixed single qubit in the chain can be made time-dependent by controlling the exchange interaction between the atom on the tip of the microscope and the target qubit in the chain.
This interaction depends on the (time-dependent) distance between the two atoms and it is equivalent to a magnetic field applied on the qubit \cite{Yan,Wieser}.
Since accurately controlling the kinematics of the STM is within the experimental reach, then a time-dependent effective magnetic field on the qubit of the chain can be generated at will by appropriately governing the time-dependent distance between the qubit of interest and the atom on the tip of the microscope \cite{Wieser}.

From a mathematical point of view, our system is characterized by the existence of $2^{N-1}$ two-dimensional dynamically invariant subspaces of the total Hilbert space $\mathcal{H}$ of the $N$-spin system.
This circumstance stems from the existence of the $2^{N-1}$ constants of motion $\hat{\sigma}_i^z\hat{\sigma}_j^z$ (with $i \neq j$) \cite{GLSM}.
In each of these two-dimensional subspaces, a basis may be chosen as: a specific appropriate state $\ket{e_k}$ of the standard basis in the Hilbert space $\mathcal{H}$ and its flipped state, that is $(\bigotimes_{k}\hat{\sigma}_k^x)\ket{e_k}$.
Then, for example, a generic subspace is spanned by the couple of states of the form $\ket{+}^{\otimes(N-m)}\ket{-}^{\otimes m}$ and $\ket{-}^{\otimes(N-m)}\ket{+}^{\otimes m}$  ($\hat{\sigma}^z \ket{\pm}= \pm 1 \ket{\pm}$).
The subspace relevant for the scope of this paper is the one involving the two states $\ket{+}^{\otimes N}$ and $\ket{-}^{\otimes N}$.
%This implies the possibility of easily generating GHZ states of the $N$-spin system through this kind of interactions between the spins, as it is well known in literature \cite{Muller,MezzacapoPRL113}.

In the light of the previous considerations, the dynamics of our $N$-spin system in each subspace can be effectively described in terms of the dynamics of a single (fictitious) two-level system.
It means that, within each two-dimensional subspace, the two involved $N$-spin states can be mapped into the two states $\ket{+}$ and $\ket{-}$ of a single generic qubit.
For the Hamiltonian model \eqref{Model}, in particular, it is remarkable to point out that all the two-dimensional subdynamics are formally governed by the same effective two-level Hamiltonian which reads
\begin{equation} \label{h}
\tilde{H} = \hbar\omega_1(t) \hat{\sigma}^z + \gamma_x \hat{\sigma}^x,
\end{equation}
where $\hat{\sigma}^x$ and $\hat{\sigma}^z$ represent the Pauli operators of the fictitious two-level system in the specific two-dimensional subspace under scrutiny.
A conspicuous consequence consists in the fact that the knowledge of exactly solvable problems of a single spin-1/2, in accordance with Eq. \eqref{h}, provides the key to exactly and simply  solve time-dependent scenarios for our $N$-spin chain.
Thus, identifying new strategies aiming at furnishing exact analytical solutions of the single-qubit dynamical problem associated to $\tilde{H}$ \cite{Barnes,MN,GdCNM} is a natural passage to make progresses toward our target.
The innovative feature of our model, then, lies in the possibility of considering appropriately engineered (time-dependent) fields, driving the transition of the $N$-spin system between two desired states belonging to a same prefixed invariant subspace, or, generally speaking, to realize the control of the time evolution of the chain from a given initial state toward a desired coherent superpositions of two states, for example.

We point out that the arguments previously reported are valid regardless the specific time-dependence of the field applied on the first spin.
This circumstance stems from the fact that the subdivision of the Hilbert space $\mathcal{H}$ into $2^{N-1}$ two-dimensional dynamically invariant subspaces, originated from the symmetries possessed by $H$, is independent of the two Hamiltonian parameters $\omega_1$ and $\gamma_x$.
In this way, it is always possible to break down the time-dependent Schr\"odinger equation for our $N$-spin system into a set of $2^{N-1}$ decoupled time-dependent Schr\"odinger equations all associated to $\tilde{H}$.

%\section{Many-Body LMSZ Transition}

%\subsection{Full Transition}
\textit{Full Transition.}
Suppose that the effective magnetic field applied on the first spin varies over time as follows
\begin{equation}
\hbar\omega_1(t)={\alpha} t/2,
\end{equation}
where $\alpha$, assumed positive without loss of generality, is related to the velocity of variation of the field, $\dot{B}_z\propto\alpha$.
We study the case in which the $N$-qubit chain is initially prepared in the state $\ket{-}^{\otimes N}$, so that the system dynamics lives within the two-dimensional subspace spanned by $\ket{-}^{\otimes N}$ and $\ket{+}^{\otimes N}$.
The situation we are considering results in a proper LMSZ scenario for the fictitious spin-1/2, effectively describing the $N$-spin chain within such a two-dimensional subspace whose Hamiltonian is given in Eq. \eqref{h}.
The fictitious two-level system, in fact, is subjected to a linearly varying magnetic field along the quantization axis ($z$ axis) and a (fictitious) constant transverse field (along the $x$ axis).
We emphasize that the actual magnetic field we are applying on the (true) ancilla qubit consists in the $z$ ramp only and that no constant transverse field is present.
The effective transverse field felt by the fictitious spin-1/2 originates from the $N$-wise interaction term between the qubits in the chain and, as shown by Eq. \eqref{h}, the relative coupling constant $\gamma_x$ determines the intensity of the (fictitious) transverse field.

It is well-known that the dynamical problem of a single spin-1/2 for an LMSZ scenario, having an arbitrary  finite duration from $t_i$ to $t_f$ (initial and final time instant, respectively), can be analytically solved \cite{Vit-Garr}.
The finite LMSZ model \cite{Vit-Garr} eliminates the nonphysical assumptions of infinite energies (both of the coupling and the detuning), being thus much closer to the experimental scenario.
%This means that we can write the exact form of the time evolution operator, solution of the Schr\"odinger equation $i\hbar\dot{U}=HU$, and then the exact form of the transition probability at any time.
We can then write the exact form of the transition probability at any time and, therefore, the time-dependent analytical expression of the transition probability of our $N$-qubit system to pass from the initial state $\ket{-}^{\otimes N}$ to the final state $\ket{+}^{\otimes N}$.
%Therefore, in this way, we get the time-dependent analytical expression of the transition probability of our $N$-qubit system to pass from the initial state $\ket{-}^{\otimes N}$ to the final state $\ket{+}^{\otimes N}$.
Of course, its asymptotic expression, when $t_f=-t_i$, in the limit of a very large duration, recovers the probability $P$ known as the LMSZ transition formula \cite{LMSZ}, that is  
\begin{equation}\label{P simple}
\begin{aligned}
P=&|^{N \otimes}\average{+|U(\infty)|-}^{\otimes N}|^2=|^{N \otimes}\average{-|U(\infty)|+}^{\otimes N}|^2= \\
=&1-\exp\{ -2\pi\gamma_x^2/\hbar\alpha \}.
\end{aligned}
\end{equation}
This expression suggests two possible well known physical scenarios of interest here, where $\alpha$ plays the role of controllable external
parameter.
In this subsection we deal with the so-called adiabatic scenario corresponding to the condition $\gamma_x \geq \hbar\alpha$. 
By applying an adiabatically driving field, when a constant (fictitious) transverse field is present, the two-level system undergoes a full transition, that is, a perfect inversion since practically $P=1$.
In the light of the mapping at the basis of the effective description of the $N$-qubit dynamics in terms of a single spin-1/2 (within the two-dimensional subspace under scrutiny), we are producing a perfect coherent inversion of all the spins at the same time.

\begin{figure}[htp] 
\begin{center}
{\subfloat[][]{\includegraphics[width=0.4\textwidth]{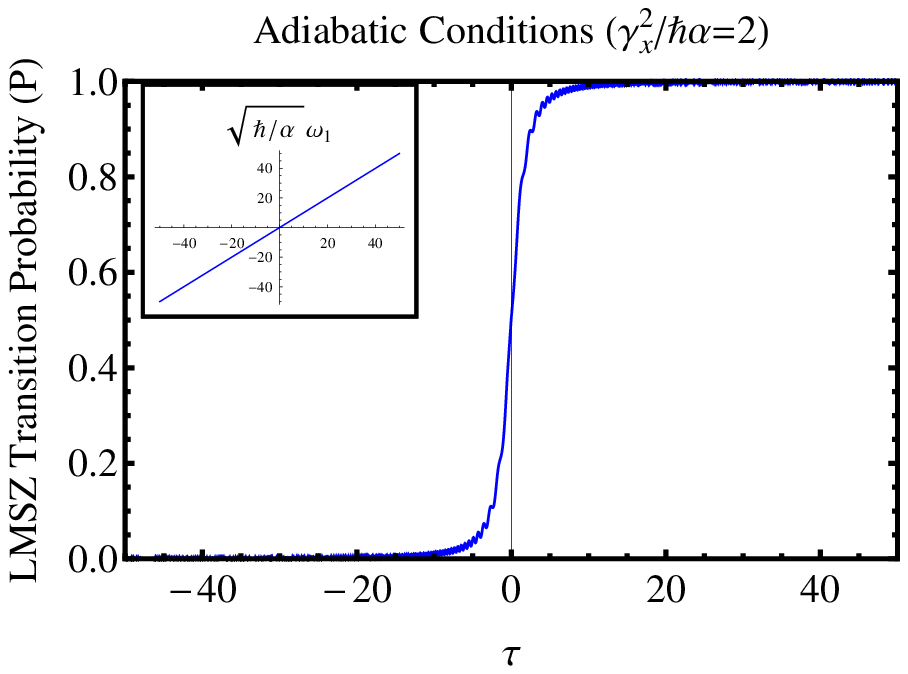}\label{fig:P1}}}
\quad
{\subfloat[][]{\includegraphics[width=0.4\textwidth]{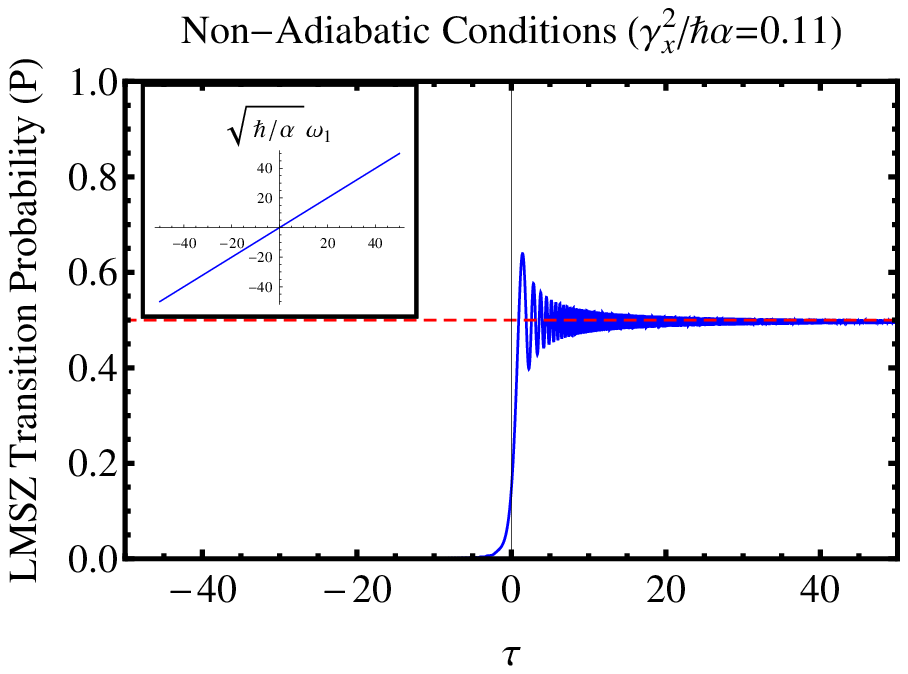}\label{fig:PhalfNA}}}
\quad
{\subfloat[][]{\includegraphics[width=0.4\textwidth]{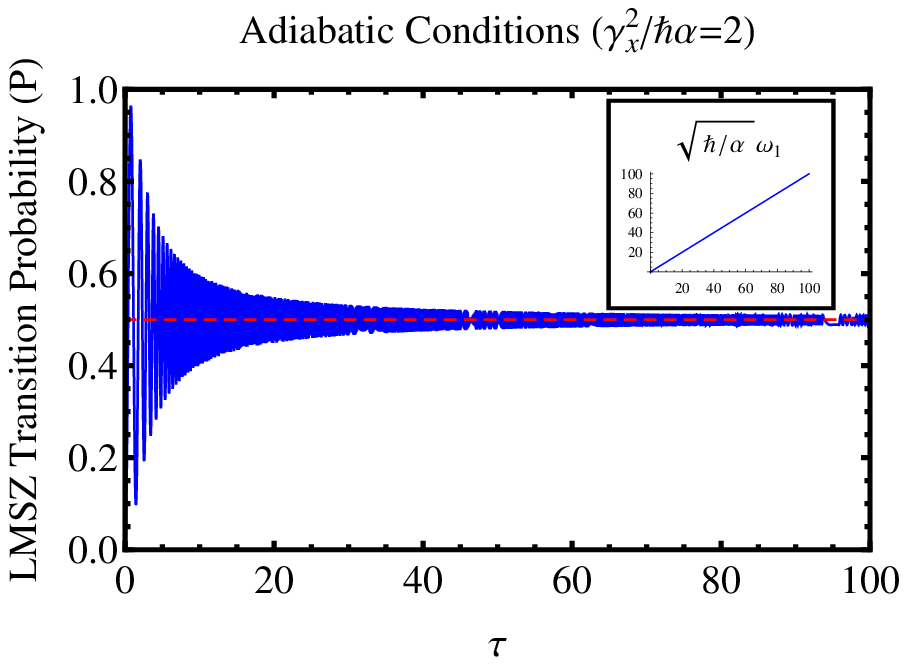}\label{fig:PhalfA}}}
\captionsetup{justification=raggedright,format=plain,skip=4pt}%
\caption{\small(Color online) Time behaviour of the LMSZ transition probability $P(\tau)$ ($\tau=\sqrt{\alpha/\hbar}~t$) of finding the $N$-qubit chain in $\ket{+}^{\otimes N}$, starting from $\ket{-}^{\otimes N}$, under: (a) the adiabatic condition $\gamma_x^2/\hbar\alpha=2$ and a symmetric ($\tau_i=-\tau_f$) ramp (inset); (b) the non-adiabatic condition $\gamma_x^2/\hbar\alpha=0.11$ [Eq. \eqref{Half Transition Condition}] and a symmetric ramp (inset); (c) the adiabatic condition $\gamma_x^2/\hbar\alpha=2$ and an asymmetric ($\tau_i=0$, $\tau_f=100$) ramp.
In the first case a full transition is performed, while in the other two cases the realization of the GHZ-state is achieved.
The constant dashed lines in the second and third plot represent $P_-^+=1/2$.}\label{fig:Plots}
\end{center}
\end{figure}

In Fig. \ref{fig:P1} the time behaviour of $P(t)$, based on the exact solution of the dynamical problem \cite{Vit-Garr}, is shown under the adiabatic condition $\gamma_x^2/\hbar\alpha=2$ as a function of the dimensionless time parameter $\tau=\sqrt{\alpha/\hbar}~t$.

%\subsection{Half Transition and Maximally Entangled GHZ States Generation}
\textit{Half Transition.}
By definition, a full transition does not realize our target.
However, since the reason lies on imposing an adiabatic control of the qubit-chain evolution, we may reasonably expect the generation of a linear superposition of $\ket{+}^{\otimes N}$ and $\ket{-}^{\otimes N}$ when the chain is instead driven under non-adiabatic conditions.
Therefore, we now explore the time evolution of the $N$-qubit system  from $t_i$ to $t_f=-t_i$ as previously considered but when $\gamma_x^2/\hbar\alpha \ll 1$, that is far from the adiabatic constraint.

It is easy to convince oneself that by setting $\gamma_x$ and $\alpha$ as follows
\begin{equation} \label{Half Transition Condition}
{\gamma_x^2 \over \hbar\alpha} = {\ln(2) \over 2\pi}  \approx 0.11,
\end{equation}
the value of $P$ from Eq. \eqref{P simple} becomes equal to $1/2$.
It means that, under such a specific condition on the parameters, the two populations $\rho_{++}=\ket{+}^{\otimes N}\bra{+}^{\otimes N}$ and $\rho_{--}=\ket{-}^{\otimes N}\bra{-}^{\otimes N}$ are asymptotically ($t \rightarrow \infty$) equal to 1/2.
Then, at very large times the $N$-qubit system can be found in the two involved states $\ket{-}^{\otimes N}$ and $\ket{+}^{\otimes N}$ with equal probability.
Being the dynamics unitary, this circumstance implies that the general state asymptotically reached by the system must be (up to a global phase factor)
\begin{equation} \label{GHZ}
\ket{\psi(\tau \rightarrow \infty)} = {\ket{+}^{\otimes N} + e^{i\varphi} \ket{-}^{\otimes N} \over \sqrt{2}}.
\end{equation}
Thus, the specific value of our control parameter, fulfilling Eq. \eqref{Half Transition Condition} for a given physical realization of our $N$-spin model, ensures an asymptotic half transition [$P(t_f \rightarrow \infty) = 1/2$], that is, allows the system to reach an equally populated superposition of the two states (see Fig. \ref{fig:PhalfNA}) involved in the subspace under scrutiny.
In Fig. \ref{fig:PhalfNA} we can see the realization of the half transition under the condition of Eq. \eqref{Half Transition Condition}.

Therefore, we can claim that a symmetric LMSZ pulse applied on the first spin of the chain for a long time interval, when the field's slope $\alpha$ and the spin-coupling $\gamma_x$ satisfy Eq. \eqref{Half Transition Condition}, is able to generate the GHZ state \eqref{GHZ}.
The relatively simple protocol outlined here, thus, is able to render all the qubits in the chain entangled in a single step only through two main ingredients: an LMSZ $\pi/2$-pulse applied on one (the first) spin only and the generation of the $N$-wise many-body interaction characterizing the model in Eq. \eqref{Model}.

The examined half transition, necessary to generate GHZ states, thus requires \textit{non-adiabatic} conditions.
%Adiabatic conditions indeed make the system to undergo a full transition $[P_-^+(t \rightarrow \infty)=1]$ from $\ket{-}^{\otimes N}$ to $\ket{+}^{\otimes N}$.
%Nevertheless, such a non-adiabatic entanglement generation strictly depends on the symmetric full level crossing (the avoided crossing happens in the middle time of the dynamics, namely $\tau=0$) until now considered.
Nevertheless, such a non-adiabatically generated entanglement strictly depends also on the symmetric ramp ($t_i=-t_f$) until now considered.
%In the `half crossing' case, instead, that is when the evolution starts exactly at the crossing, the LMSZ transition probability reads $P=[1-\exp(-2\pi\gamma_x^2/\hbar\alpha)]/2$ \cite{Vit-Garr,Vit-Zlat}.
In the asymmetric case ($t_i=0$, $t_f=100$), instead, the LMSZ transition probability reads \cite{Vit-Garr,Vit-Zlat}
\begin{equation}\label{asymm-ramp prob}
P={1-\exp\{-\pi\gamma_x^2/2\hbar\alpha\} \over 2}.
\end{equation}
%Thus, the half-crossing dynamics can induce a half transition, with the consequent GHZ state generation for the qubit chain under \textit{adiabatic} conditions.
Thus, the asymmetric ramp can induce a half transition, with the consequent GHZ state generation for the qubit chain, under \textit{adiabatic} conditions.
In Fig. \ref{fig:PhalfA} such an effect is clearly shown by plotting the exact time dependence of the LMSZ transition probability for $\gamma_x^2/\hbar\alpha=2$ and the field turned of at $\tau=0$ (inset of Fig. \ref{fig:PhalfA}).
This aspect highlights the physical relevance of the half-crossing dynamics, besides opening possible interesting applications for $N$-qubit scenarios.

It deserves to be stressed that the rapid oscillations seen in the plots of Fig. \ref{fig:Plots} are artifacts of the LMSZ model.
Assuming a slight modification of the Allen-Eberly-Hioe model \cite{Vit-Garr}, namely a constant coupling and a tangent-detuning (or ramp), it is possible to see that the transition probability, in this instance, presents no such oscillations \cite{Vit-Garr}.

We underline that the analogous dynamics (full and half transitions) is obtained within each dynamically invariant two-dimensional subspace when the $N$ qubits start from another standard basis state $\ket{e_k}$.
In this case, we asymptotically get a GHZ-like state of the form
\begin{equation}
\ket{\psi(\tau \rightarrow \infty)} = { \ket{e_k} + e^{i\varphi} [(\bigotimes_{k}\hat{\sigma}_k^x)\ket{e_k}] \over \sqrt{2} }.
\end{equation}
Furthermore, it is important to stress that analogous results can be produced by considering a generalization of the model by introducing in the Hamiltonian model \eqref{Model} a further $N$-wise interaction term, namely $\bigotimes_{k=1}^N \hat{\sigma}_k^y$.
In this instance too, the symmetry possessed by the Hamiltonian determines the existence of $2^{N-1}$ dynamically invariant subspaces in the Hilbert space of the $N$-spin system \cite{GLSM}.
Then, also in this case we can reduce the $N$-qubit dynamical problem to $2^{N-1}$ effective two-level dynamical problems \cite{GLSM}.
Although a slight modification of the LMSZ transition probability would occur, the latter remains analytically derivable \cite{GNMV} and the possibility of GHZ-state generation, as reported in this paper, is preserved. 
The interaction $\bigotimes_{k=1}^N \hat{\sigma}_k^z$, instead, would contribute with a constant term in each two-dimensional subdynamics, not affecting thus the LMSZ transition probability.

It is now important to discuss the possible time scales of GHZ-state generation for the procedure outlined here.
As suggested by Eq. \eqref{P simple}, the time scale depends on both the magnetic field gradient and the coupling parameter.
In a microfabricated ion trap the magnetic field gradient can reach values as large as 150-200 T/m \cite{Weidt}.
As the spin-spin coupling is concerned, for Rydberg atoms and ions it can reach few MHz \cite{Urban,Gaetan,MullerZoller}, implying GHZ-state generation on the sub-microsecond scale.
For microwave-driven trapped ions, instead, the effective spin-spin coupling, proportional to the magnetic-field gradient, can reach the kHz range \cite{Weidt,Johanning}.
Finally, in case of two-coupled qubits \cite{GNMV,GVM1} or qutrits \cite{GVM2}, in nuclear magnetic resonance the typical range of spin-spin coupling is 10-300 Hz (depending on the molecule \cite{Chuang}), with a consequent GHZ(Bell)-state generation on the millisecond scale.

\textit{Conclusions.}
In this work we have shown that, thanks to the unconventional $N$-wise interaction generated in a $N$-spin-qubit chain, GHZ states and, more generally, multipartite entangled states can be easily produced through non-local control, namely by applying a single LMSZ $\pi/2$-pulse on just one (ancilla) qubit.
Thanks to the exact solution at our disposal, our technique results clear and relatively simple to apply.
We brought to light that the generation of GHZ states can be achieved under both \textit{adiabatic} and \textit{non-adiabatic} conditions, depending on controllable characteristics of the LMSZ ramp.

Moreover, we emphasize that, due to the peculiar non-classical properties of the GHZ state, it would be of remarkable interest to investigate the applicability of the LMSZ-based protocol here reported on other types of coupled systems \cite{GMMM}, especially in presence of environmental noise by  taking advantage of both the non-Hermitian \cite{Brody,GMSVF} and the Wigner approach \cite{Kapral,SHGM}.
Finally, we remark that our results can be read and reinterpreted in terms of fermion variables on the basis of the correspondence between $N$-wise spin models and many-body fermion models \cite{Casanova}.

%\section{Acknowledgements}
\textit{Acknowledgements.}
RG and DV acknowledge financial support from the PRIN Project PRJ-0232 - Impact of Climate Change on the biogeochemistry of Contaminants in the Mediterranean sea (ICCC).

\end{document}